\documentclass[onecollarge]{svjour2}       
\usepackage{amsfonts}

\usepackage{graphicx}
\usepackage{cite}
\usepackage{latexsym}

\usepackage[USenglish]{babel} 
\usepackage[T1]{fontenc}
\usepackage[ansinew]{inputenc}

\usepackage{lmodern} 

\usepackage{graphicx} 
\begin{document}
\title{Possible visualization of a superfluid vortex loop attached to an oscillating beam.}


\titlerunning{Visualization of vortex loop in superfluid.}        

\author{E. Zemma   \and M. Tsubota    \and J. Luzuriaga}


\institute{E. Zemma  \at
              Centro At\'omico Bariloche, (8400) S.C. Bariloche, CNEA, Inst. Balseiro UNC,CONICET, Argentina\\
                           \email{zemma@cab.cnea.gov.ar}           
\and  
  M. Tsubota  \at  
         Department of Physics, Osaka City University, Osaka 558-8585, Japan 
          \and
           J. Luzuriaga \at
              Centro At\'omico Bariloche,(8400)S.C. Bariloche, CNEA, Inst. Balseiro,UNC, Argentina
}

\date{Received: date / Accepted: date}

\maketitle









\begin{abstract}
Visualization using tracer particles is a relatively new tool available for the study of
superfluid turbulence and flow, which is applied here to oscillating objects submerged
in the liquid.
We report observations of a structure  seen in  videos taken from outside a cryostat filled
with superfluid helium at 2 K, which is possibly a vortex loop attached to an oscillator. The feature, which has the shape of an incomplete arch, is visualized due to the presence of  solid $H_2$ tracer particles and is  attached to a beam oscillating at 38 Hz in the liquid. It has been recorded in videos taken at 240 frames per second (FPS), fast enough to take $\sim $ 6 images per period. This makes it possible to follow the  structure,  and to see that  is not rigid. It   moves with respect to the oscillator, and its displacement is in phase with the velocity of the 
moving beam. Analyzing the motion, we come to the conclusion that we may be observing a superfluid
vortex attached to the beam and decorated by the hydrogen particles. An alternative model, considering 
a solid hydrogen filament, has also been analyzed, but the observed phase between the movement
of the beam and the filamentary structure is better explained by  the superfluid vortex hypothesis.


\end{abstract}

\maketitle

\section{Introduction}

In the superfluid phase of liquid Helium, below 2.177 K,  the circulation is  quantized
in units of a flux quantum ($ \kappa = h/m_{He}  \simeq 10^{-3} cm^2/s $ where $m_{He}$ is the mass of a He$^4$ atom  and $h$ is the Planck constant). The existence of vortices with a single flux quantum was  independently proposed by Feynman and Onsager \cite{feynman1957superfluidity,onsager1949suppl} and the first measurements showing quantized circulation were made by Vinen\cite{vinen1961detection}. More recently  superfluid vortices have been observed by Bewley et al\cite{bewley2006}, and  this group has developed solid hydrogen tracers\cite{bewley2009generation,bewley2008particles} to visualize the flow and has observed many interesting features of vortex physics such as re-connections\cite{bewley2006,bewley2009generation} and  Kelvin waves \cite{fonda2014direct}. Visualizations of turbulence generated by counterflow have also been obtained by this 
technique\cite{LaMantiaSkrbek2014,la2013velocity,paoletti2008velocity} which is becoming a 
powerful tool in the study of 
Quantum Turbulence\cite{guo2014visualization}. 

However, apart from some preliminary work\cite{zemma2013part1}, the visualization of flow 
around objects oscillating in superfluids has not been explored 
so much, although 
 Vinen and Skrbek\cite{vinen2010quantum,vinen2014quantum}  have pointed out that   tracer imaging of superfluid oscillatory flows with a classical analogue could provide valuable information. We have recently developed a simple system using solid hydrogen particles for  visualizing flow around oscillating objects in superfluid helium\cite{zemma2013part1}.
 In the following we present further  observations made using this system which show a behavior which is consistent with the presence of a superfluid vortex half-loop attached to a beam  oscillating in liquid
helium.  The loop is observed to expand and contract and is attached to the beam throughout our observation.
 To our knowledge, such behavior has not been directly visualized previously, although vortex loops have been observed indirectly through their attachment and detachment\cite{GotoTsubota2008} and  discussed theoretically before\cite{Hanninen2009,godfrey2001new,godfrey2000stable}. We therefore believe that 
our observations could shed new light on the problem of superfluid turbulence, in particular on 
the behavior of vortices attached to a solid boundary with oscillating flow and their stability and
pinning.
    
\section{Experimental Details}
\begin{figure}
\includegraphics[width=0.7\textwidth, angle=0]{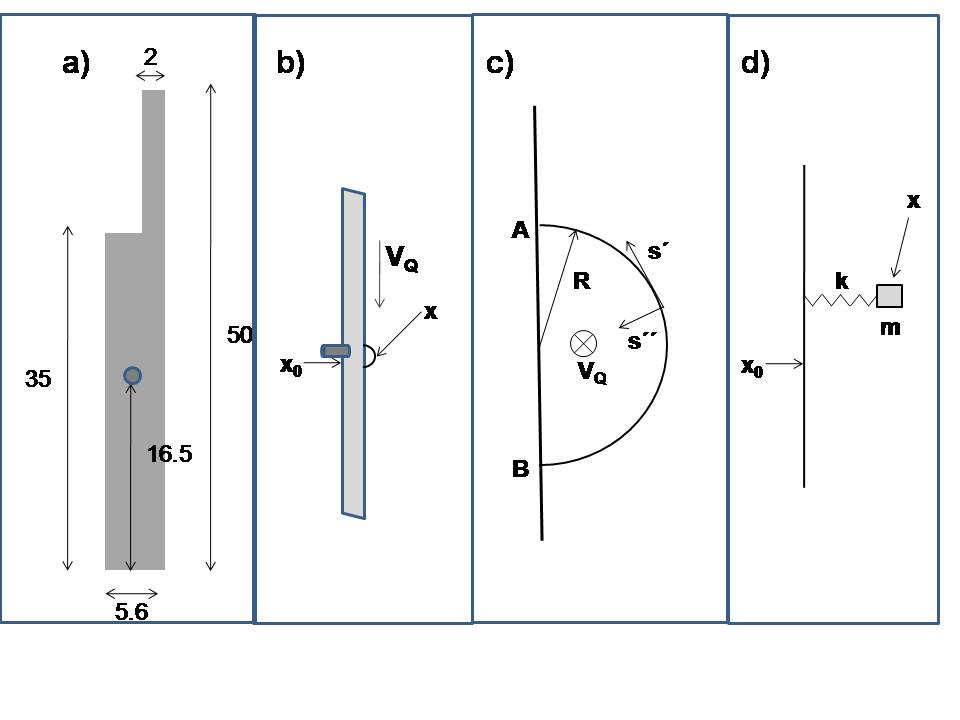}
\caption{a)Diagram of the vibrating beam, with measurements in millimeters. 
The position of the driving magnet is shown as a circle in this
frontal view of the setup.
b) Position of the points $x$ and $x_0$ (see text) in the approximate
perspective seen in the videos. $V_Q$ represents the local superfluid velocity
due to the heat load of the cryostat. The driving magnet and its position are shown as a 
cylinder in perspective.  
c) Proposed geometry of the vortex loop and definition of some quantities used in the
analysis
d) Alternative mass-spring model considering a possible hydrogen filament, instead of the 
loop shown in c). }
\label{diagrama}
\end{figure}
The experimental arrangement has been described in detail elsewhere\cite{zemma2013part1}. The system studied consists of a vibrating beam, driven magnetically by using a permanent magnet attached to the beam and a coil fixed to a rigid frame. In this way the beam  oscillates with velocity perpendicular to its wide dimension at a frequency of 38 Hz. A sketch of the geometry is found in Fig. \ref{diagrama}. Videos are taken with a camera at 240 frames per second (FPS) so the time interval is 4.17 ms between frames and we use this as our
time reference to calculate velocities. The helium temperature   was  2.07 K throughout the experiment.

A mixture of one part hydrogen to 50 parts  helium gas at 500 torr is introduced from room temperature to form the solid hydrogen particles and around a hundred cubic centimeters of gas are injected each time. 
To illuminate the tracer particles we used a green laser beam. The frozen
H$_2$ particles are not expected to absorb significant energy in the visible \cite{CaloryPart}.
The laser is on the outside of the dewar and the light passes through
an optical fiber which ends less than a centimeter away from the oscillating beam, illuminating the particles perpendicular to the line of sight of the
camera. The fiber is polished at the end, giving a three dimensional cone of
light.  Distances in the image are calibrated
with respect to the measured dimensions of the small magnet (a cylinder 5mm long
and 3 mm in diameter), which is used to drive the oscillator.  The size of a pixel
corresponds to around 70 microns in the object but the light could be scattered
from particles which are smaller than this. It is hard to evaluate the minimum
observable dimension, but we estimate our particles to be distributed  in size from well
below 70 microns to 200 microns.

An important feature of our setup is that we are forced to remove the outer nitrogen Dewar
to avoid the blurring of the images produced by the boiling  nitrogen. For this reason, the  
heat load is considerable. We can estimate it by measuring the volume of He evaporated as
a function of time and using the known latent heat of evaporation. The heat input $ \dot{Q}$ is not constant, but has been roughly measured to lie between 800 and 470 mW. The  helium Dewar is 6 cm in diameter, so the calculated 
counterflow velocity $v_Q$ is between 0.085 and 0.05 cm/sec   if we assume a uniform heat flow. These numbers are only rough estimates since the geometry is not simple, the heat input comes from radiation through the walls,
conduction down the glass walls, etc. On  the Dewar there is a flange of 5 cm diameter  about 5 cm above the vibrating beam which also complicates the counterflow geometry.

\section{Experimental Results}

\begin{figure}[ht]
\includegraphics[width=0.8\textwidth, angle=0]{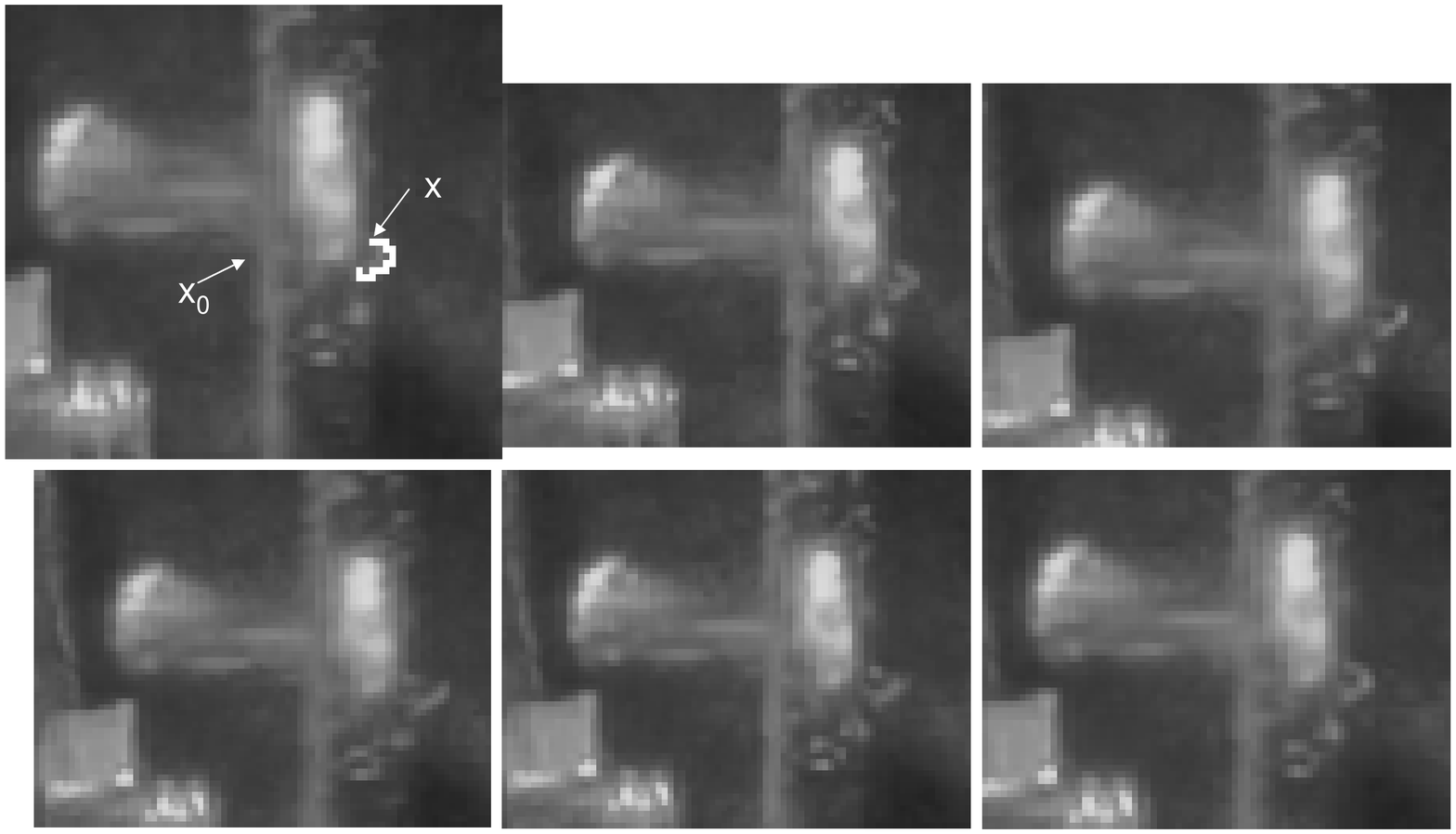} 
\caption{Still images of a video. 
The six images correspond to successive frames, taken at 4.17 ms intervals
and therefore covering a complete cycle of the oscillator whose frequency
$\omega_B / 2\pi = 38 $Hz. The pixel size can be seen at this magnification.   } 
\label{fotos}
\end{figure}

The main observation of the experiment is the existence of a structure  
formed by the H$_2$ particles 
that seems to be attached to the beam 
and oscillates with it, though not in a rigid fashion. It seems to elongate and contract 
when the beam oscillates, and has a curved shape, somewhat resembling an incomplete arch.
This structure is seen in all videos taken during the experimental run. 
We show still images in Fig. \ref{fotos} taken over one complete cycle of the oscillator. The 
images are amplified close to the maximum resolution and the pixel structure can be clearly seen.

For  analysis, we have chosen to follow two points marked in Fig. \ref{fotos} and Fig. \ref{diagrama} 
as $x$ and $x_0$. Point
$x_0$  is fixed to the oscillating beam and $x$ is a point on the arch, which is seen to shrink and grow
periodically. We have followed  $x$ and $x_0$ over several cycles of the oscillator, observing the 
images by eye, and digitizing the positions $x$ and $x_0$ by means of 
a computer. The difference in the position of these two points, which have a sinusoidal motion, is
proportional to the length of the arch formed by the decorating particles. We have taken a well 
defined and easy to follow point for $x_0$ instead of the base of the arch, so $x-x_0 $ has a
constant displacement superposed to the periodic component but the sinusoidal variation
is proportional to the length of the arch.   

\begin{figure}[ht]
\includegraphics[width=0.8\textwidth, angle=0]{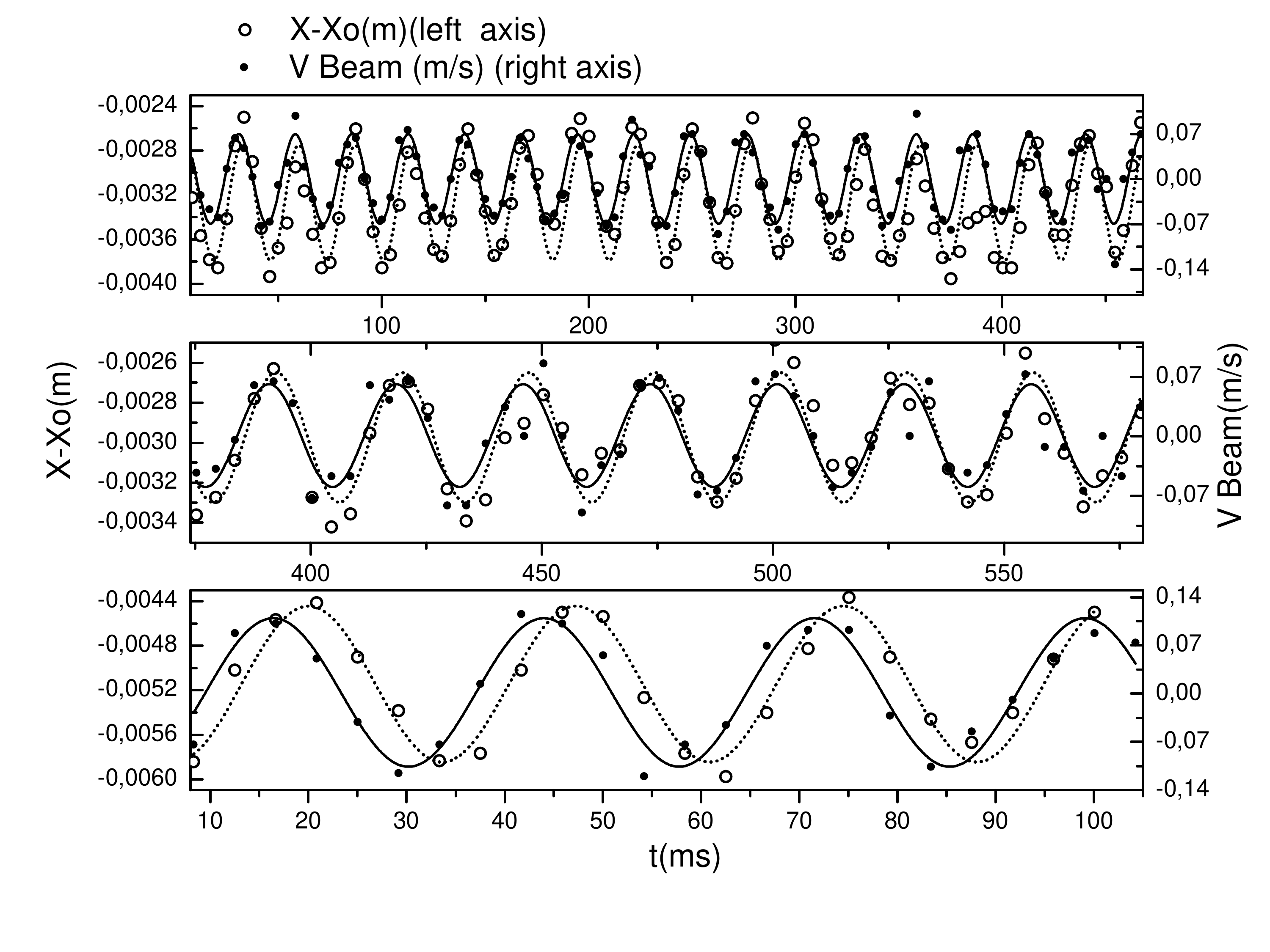}
\caption{Comparisons of beam velocity $V_{Beam} $ and $ x-x_0$ for three sections
of the videos. Open  circles correspond to the velocity of the beam 
  and filled  circles to the distance $x-x_0$. The dashed and full lines are sinusoidal fits
	to the data and it can be seen from the graphs  that  $V_{Beam} $ and $ x-x_0$
are almost in phase.}
\label{oscilacion}
\end{figure}

Digitizing the positions of $x$ and $x_0$ we are able to calculate the velocities of the beam 
and the arch as a function
of time (Using  the $x_0$ positions in successive images and dividing by the known time of 4.17 ms between frames). We  can also evaluate the distance $(x-x_0)$ in each frame.  
The results are shown on Fig. \ref{oscilacion}. Open circles correspond to the velocity of the beam   and filled circles to the distance $x-x_0$. A few cycles are shown, and they include information from three different video sections, all taken the same day but at different times. 
We have fitted the experimental points with sinusoidal functions by least squares, and the
results are shown as  lines in the plots.  The fits show clearly that the velocity of the beam is in phase with the relative displacement  $x-x_0$. Bearing in mind that  $x-x_0$
is proportional to the length of the arch formed by the solid particles, the length of this structure therefore is in phase with the velocity of the beam. We have also evaluated the velocities of $x$ and $x_0$ separately, and find that they are around
90 degrees out of phase with each other. Therefore  the arch structure is not rigidly attached to the beam, but  shows an internal shrinking and growing movement in phase with the velocity.

\begin{figure}[ht]
\includegraphics[width=0.8\textwidth, angle=0]{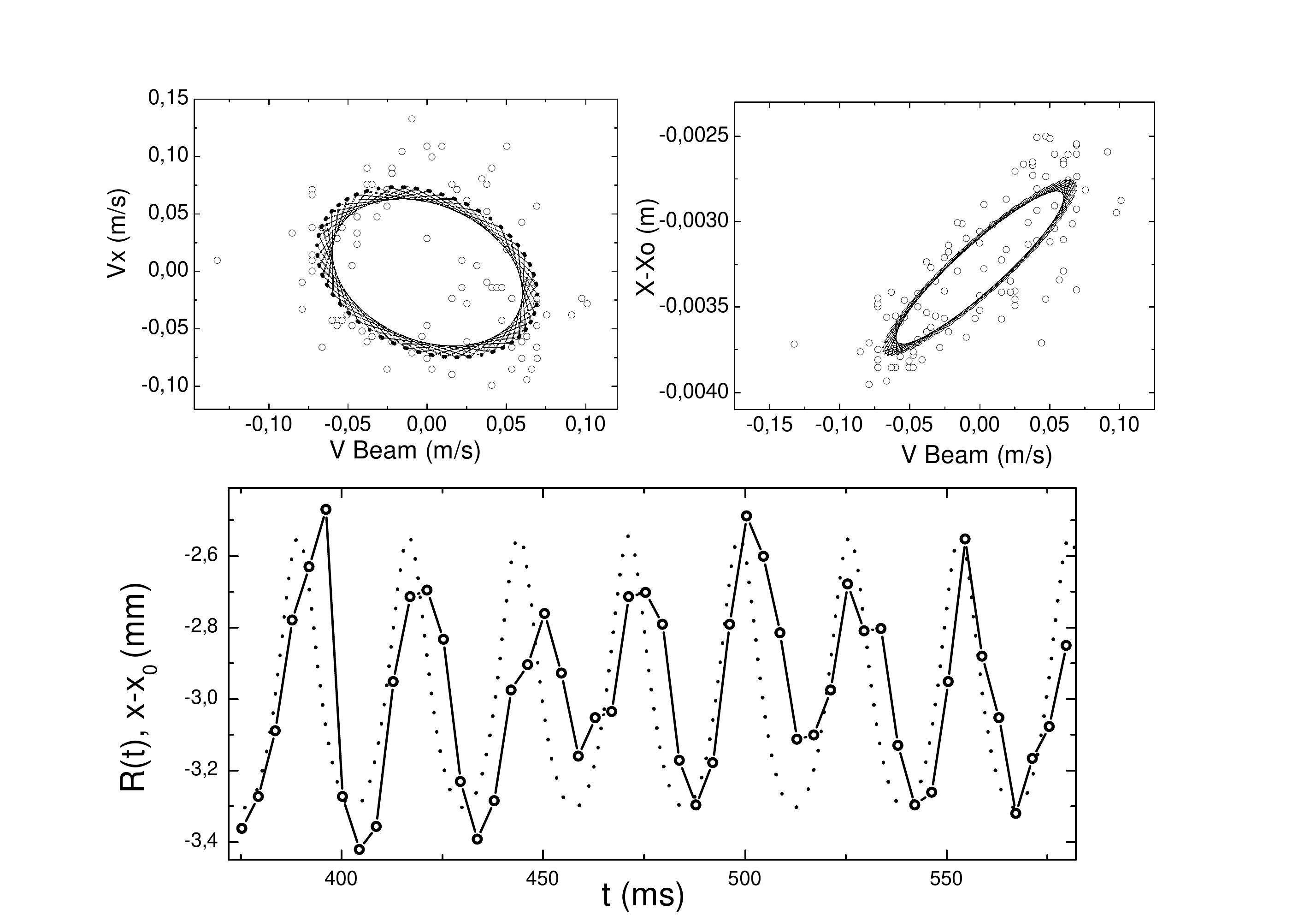}
\caption{Upper panel: Parametric plots of $ Vx $ \textit {vs} $V_{Beam} $  (left), and $x-x_0$  \textit {vs} $V_{Beam} $ (right). Open circles: experimental points, lines: sinusoidal fits to the data, 
equivalent to the full and dotted lines shown in Fig. 3. 
The almost circular plot on the left indicates that the corresponding sinusoidal motions 
are 90 degrees out of phase, while the highly tilted ellipse at the right implies that 
the oscillations are almost in phase. 
Lower panel: Fit of $x-x_0 \propto R(t) $ using Eq. \ref{eqcuatro} and fitting parameters $A= 0.37 $ , $ B= 0.001$. Lines and open circles correspond to the measured $x-x_0$ and the fit is shown in
dotted lines. The good agreement  indicates that the vortex loop model gives an adequate description of
the observations. }
\label{fases}
\end{figure}

The phase difference between velocities and $ x-x_0$ is shown in a different way in Fig. \ref{fases}. 
We show the data in a parametric plot, on the left hand side the velocity of the 
the beam ($V_{Beam}$) is plotted against a point corresponding to the top of the arch
($Vx$)   and it is seen that the plot is 
almost circular, as corresponds to a parametric representation of two sinusoidal functions with a 90 degree phase difference, while
on the right a plot of $x-x_0$ against $V_{Beam}$ shows a very elongated ellipse at a  45 degree angle, 
as would be seen for two sinusoidal functions that are almost in phase.

Although we only show a few periods, videos are longer than this, and the motion of the arch  was observed to remain basically unchanged throughout the whole experimental run. 
 The video
camera was not always on, but the oscillation of the beam was not changed  for 13 minutes. During this
time  we filmed ten sections of video at 240 and 480 frames per second, and in all
of them we see the loop, with similar behavior. The sections shown are representative of
the first and last parts filmed, at 240 FPS. We have not included data for the 480 FPS because 
the  resolution and lighting are not good enough for measuring quantitatively, although
the images show movement of the loop that is compatible with that seen at 240 FPS.
Several experiments with the same setup were performed, but the structure seen here was seen 
in only in this particular run when the videos were  analyzed in detail later.
 It seems therefore that the formation of a loop is not a reproducible 
feature, but depends on several uncontrolled factors as is expected in turbulent regimes.

\section{Discussion}

We have come to the conclusion that   the behavior observed is consistent with the motion expected from a  vortex half-loop pinned to the 
beam. This appears to cover the main facts,  although other possibilities have also been considered.
We use for our analysis the relationship obtained by 
Schwarz\cite{schwarz1985} for a stable vortex loop. According to his\cite{schwarz1985} Eq. 20 
 a vortex loop of radius $R_0$ will not shrink or 
grow if it moves with a velocity $v_s$ with respect to the superfluid:
\begin{equation}
v_s = \frac{\kappa}{4\pi R_0} \ln \left[\frac{8R_0}{e^{1/4} a_0} \right]
\label{eqdos} 
\end{equation}
with $a_0 \approx 1.3 \times 10^{-8}$ cm an adjustable parameter roughly equivalent to the size of the vortex core. Conversely, if the loop is fixed in position, a flow of the
superfluid with velocity  $v_s$  maintains a stable radius, if the sign of the velocity and 
  the vorticity of the loop are in the right orientation.
 Section IV A of Ref. \cite{schwarz1985}  is also relevant to our situation, 
since it discusses a bent vortex loop pinned at two points and his Fig. 29 shows the 
shape calculated for different values of superfluid velocity.
In our case, using Eq. \ref{eqdos} and an average curvature radius of 0.3 mm the velocity $v_s$ for a stable half loop would be 0.027 cm/s. 
The estimate for the average counterflow velocity $V_Q$ is  a factor between 2 to 3 times  greater  than $v_s$. However the flow due to $V_Q$  close to the beam  
 would probably  be  lower than  $V_Q$ itself  due to the obstacles present.
If we assume that $V_Q$  along the beam is modulated by the periodic motion of the beam, we could explain the 
stretching and contraction of the arch attached to the oscillating beam. The modulation due 
to the beam $V_{Beam}$ and the counterflow velocity would add to produce a time dependent
velocity
 \begin{equation}
v_{Osc}(t)= A \cdot V_Q + B \cdot V_{Beam} \sin(\omega t )
\label{eqtres}
\end{equation}
where $A$ and $B$ are adjustable parameters, $V_{Beam}$ is the measured velocity of the beam,
and $\omega $ the oscillating frequency.

From Eq. \ref{eqdos} we can obtain an approximate  expression for the radius of the loop due to the 
time dependent velocity
\begin{equation}
R (t)=  \frac{\kappa}{4\pi (A \cdot V_Q + B \cdot V_{Beam} \sin(\omega t ))} \cdot C
\label{eqcuatro}
\end{equation}
Where $R_0 = \frac{\kappa}{4 \pi A \cdot V_Q } \cdot C \approx 0.3 $ mm would be the stable radius if $V_Q$ has no modulation, and we 
have not considered significant   the modulation of the logarithmic term, including it as a constant 
$C =  \ln \left[\frac{8R_0}{e^{1/4} a_0} \right]$. We have fitted $(x-x_0) \sim R(t)$ to the expression given by Eq. \ref{eqcuatro} and we show the results in the bottom panel of Fig. \ref{fases}. The adjustable parameters used are $A= 0.37 $ and $B= 0.001 $, and a shift has been introduced to compensate for the  arbitrary origin when choosing $x_0$. In this case we have used the lower value of 
$V_Q = 0.05 $ cm/s. The fit is quite good, and although it is not a least squares fit and the 
parameters were chosen by hand, the equation is capable  of reproducing the  observations.

A second possibility is that the structure seen is not a vortex loop, but a filament of solid 
hydrogen, as has been observed by Gordon \textit{et al}\cite{gordon2007filament,gordon2009catalysis}. 
We could model a solid
filament as a mass attached to a spring, as is shown in Fig. \ref{diagrama} d). The movement of the
attaching point $x_0$ would move the mass at $x$ as a forced harmonic oscillator
\begin{equation}
m\ddot{x} + \gamma \dot{x}+ k (x-x_0) = 0
\label{eqcinco}
\end{equation}
the well known solution for this equation is  harmonic motion, with a  phase difference
$ \phi $ between $x_0$ and $x$. Our observed phase difference is of around 90 degrees,
between these quantities, which would be the case only if the driving frequency 
$ \omega _B / 2 \pi $ of 38 Hz
were accidentally  close to the resonance frequency of the filament $\omega _f =\sqrt{k/m}$. It is highly unlikely that $\omega _f$
is the same as the frequency of the beam $\omega _B$, and it could  be expected 
that $ \phi $ would either be zero (if $\omega _f >> \omega _B $) or 180 degrees 
(if $\omega _f << \omega _B $). 
A third possibility is that the oscillating filament could be overdamped by the influence of the
normal component, but in this case $x$ would follow the fluid, which in the 
reference frame of the laboratory, moves  180 degrees out of fase with the point $x_0$ 
belonging to the beam. 
In fact, our observation could be a combination of a hydrogen filament and part of
a vortex loop, attached to the end of the filament and a point in  the beam.
 The filament shown in Fig. 3 of 
Ref. \cite{gordon2009catalysis} shows movement in the normal fluid, but to have the 
phase observed here some form of vortex section, closing the loop and with movement 
given by Eq. \ref{eqcuatro} seems to be necessary to explain the observed behavior.

For  further analysis, the local approximation could also be used, and in this approach  
the dynamics of a quantized vortex is described by the equation proposed by Schwarz\cite{schwarz1988three}
\begin{equation}
\frac{d\textbf{s}}{dt} = \beta \textbf{s} ^\prime \times \textbf{s}^{ \prime \prime} + \textbf{v}_s + \alpha  \textbf{s} ^\prime \times ( \textbf{v}_n - \textbf{v}_s -\beta \textbf{s} ^\prime \times \textbf{s}^{ \prime \prime})
\label{LocalAp} 
\end{equation}
here $\textbf{s}$ is a point on the core of a vortex loop, $\beta$, $\alpha $ the coefficient of mutual friction, $ \textbf{v}_n $ and  $\textbf{v}_s$ the velocities of the normal and superfluid fractions, 
 $\textbf{s}^{ \prime } $ is the tangent to the vortex core,  $\textbf{s}^{ \prime \prime}$ the principal radius. 
The proposed structure of the  vortex half-loop and the definition of the vector 
quantities $\textbf{s}^{ \prime } $ and $\textbf{s}^{ \prime \prime}$ in Eq. \ref{LocalAp} are shown in Fig. \ref{diagrama}.
We can assume that the vibrating beam pushes both the normal fluid and the superfluid together,
as well as modulating $V_Q$ as described above, so we have 
\begin{equation}
\textbf{v}_n \cdot \hat{\textbf{r}} = \textbf{v}_s \cdot \hat{\textbf{r}} = v_{1} \sin \omega t \cdot \cos \varphi
\label{vosc} 
\end{equation}
with $\varphi $ the angle between the local velocity and the vortex. Then Eq. \ref{LocalAp} implies 
\begin{equation}
\frac{dR}{dt} =  D \cdot v_{1} \sin \omega t \cdot \cos \varphi - \frac{\alpha \beta }{R} 
\label{dRdt} 
\end{equation}
The equation is local, so that $\frac{dR}{dt}$ changes over the circumference with $\cos \varphi $.
It also changes the shape of the half loop, depending 
on the relative orientation of loop and velocity, 
but we can get an approximate value for the average radius $ \left\langle R \right\rangle$
 neglecting  the second term with respect to  the first  and integrating in time and over $\varphi $
\begin{equation} 
\left\langle R \right\rangle= R_0 -  D \frac{v_{1}}{\omega} \cos \omega t.
\label{Rmean} 
\end{equation}
with $D$ a parameter taking into account the angular integration over $\varphi $. 
 We do not have enough resolution to 
detect the changes of shape implied by Eq. \ref{Rmean}, but it would appear that the  effect  is 
smaller than that of Eq. \ref{eqcuatro}. Furthermore it gives  motion with a phase that is at
 90 degrees from the velocity of the beam, instead of the in phase motion observed.

In fact, Eq. \ref{eqcuatro} can be taken as a quasi static non local 
solution including a modulation of $V_Q$, and Eq. \ref{Rmean} as a local time dependent correction 
due to the presence of $v_{1}$ whose importance is given by the parameter $D$. The good
fit obtained with Eq. \ref{eqcuatro}, seen in Fig. \ref{fases} seems to indicate that the
effect of the correction of  Eq. \ref{Rmean}  is not very large, although it could be responsible for the
fact that in the measurements $x-x_0$ and $V_{Beam}$ are  not always in phase. In fact, seen
over many cycles, there are small irregularities in the motion, where the structure seems to 
stretch more or less. However, the overall stability is preserved, as stated earlier, over
at least the 13 minutes where we have partial observations. Since the frequency is 38 Hz, the number of
cycles over which the stretching and shrinking  is repeated  is of order  $ 3 \times 10^4$.  

In conclusion, we have observed, by decoration with solid $H_2$ tracers, a structure which 
moves attached to a vibrating beam. From an analysis of  possible models for the 
observed motion, we conclude 
that it behaves as expected for a vortex half-loop attached
to the oscillator. This accounts for the phase relationship  between position and
velocity, and we obtain  a good fit between the model and the video images, while
an  alternative explanation postulating a hydrogen filament requires an unlikely coincidence 
between the driving frequency and the natural frequency of the hydrogen filament. 
A third model, considering that the motion is due to  the  drag of the normal component 
on an over damped filament, would also produce a phase difference  that is not  the one  observed. 

\begin{acknowledgements}
This work was partially supported by  06/C432  grant from U.N. Cuyo and 
CONICET-Czech Academy of Sciences Scientific Cooperation agreement.  
\end{acknowledgements}

\providecommand{\newblock}{}


\begin{thebibliography}{10}
\expandafter\ifx\csname url\endcsname\relax
  \def\url#1{{\tt #1}}\fi
\expandafter\ifx\csname urlprefix\endcsname\relax\def\urlprefix{URL }\fi
\providecommand{\eprint}[2][]{\url{#2}}

\bibitem{feynman1957superfluidity}
Feynman R~P 1957 {\em Reviews of modern physics\/} {\bf 29} 205--212

\bibitem{onsager1949suppl}
Onsager L 1949 {\em Nuovo cimento\/} {\bf 6} 249--250

\bibitem{vinen1961detection}
Vinen W 1961 {\em Proceedings of the Royal Society of London. Series A.
  Mathematical and Physical Sciences\/} {\bf 260} 218--236

\bibitem{bewley2006}
Bewley G, Lathrop D and Sreenivasan K 2006 {\em Nature\/} {\bf 441} 588--588

\bibitem{bewley2009generation}
Bewley G 2009 {\em Cryogenics\/} {\bf 49} 549--553

\bibitem{bewley2008particles}
Bewley G, Sreenivasan K and Lathrop D 2008 {\em Experiments in Fluids\/} {\bf
  44} 887--896

\bibitem{fonda2014direct}
Fonda E, Meichle D~P, Ouellette N~T, Hormoz S and Lathrop D~P 2014 {\em
  Proceedings of the National Academy of Sciences\/} {\bf 111} 4707--4710

\bibitem{LaMantiaSkrbek2014}
La~Mantia M and Skrbek L 2014 {\em EPL (Europhysics Letters)\/} {\bf 105} 46002

\bibitem{la2013velocity}
La~Mantia M, Duda D, Rotter M and Skrbek L 2013 {\em Procedia IUTAM\/} {\bf 9}
  79--85

\bibitem{paoletti2008velocity}
Paoletti M, Fisher M, Sreenivasan K and Lathrop D 2008 {\em Physical review
  letters\/} {\bf 101} 154501

\bibitem{guo2014visualization}
Guo W, La~Mantia M, Lathrop D~P and Van~Sciver S~W 2014 {\em Proceedings of the
  National Academy of Sciences\/} {\bf 111} 4653--4658

\bibitem{zemma2013part1}
Zemma E and Luzuriaga J 2013 {\em Journal of Low Temperature Physics\/} {\bf
  173} 71--79

\bibitem{vinen2010quantum}
Vinen W 2010 {\em Journal of Low Temperature Physics\/} {\bf 161} 419--444

\bibitem{vinen2014quantum}
Vinen W~F and Skrbek L 2014 {\em Proceedings of the National Academy of
  Sciences\/} {\bf 111} 4699--4706

\bibitem{GotoTsubota2008}
Goto R, Fujiyama S, Yano H, Nago Y, Hashimoto N, Obara K, Ishikawa O, Tsubota M
  and Hata T 2008 {\em Phys. Rev. Lett.\/} {\bf 100}(4) 045301

\bibitem{Hanninen2009}
H\"anninen R, Tsubota M and Vinen W~F 2007 {\em Phys. Rev. B\/} {\bf 75}(6)
  064502

\bibitem{godfrey2001new}
Godfrey S and Samuels D 2001 {\em Journal of low temperature physics\/} {\bf
  125} 69--85

\bibitem{godfrey2000stable}
Godfrey S~P and Samuels D~C 2000 {\em Physical Review B\/} {\bf 61} 4190

\bibitem{CaloryPart}
Penney R and Hunt T~K 1968 {\em Phys. Rev.\/} {\bf 169}(1) 228--228

\bibitem{schwarz1985}
Schwarz K 1985 {\em Physical Review B\/} {\bf 31} 5782

\bibitem{gordon2007filament}
Gordon E~B, Nishida R, Nomura R and Okuda Y 2007 {\em JETP Letters\/} {\bf 85}
  581--584

\bibitem{gordon2009catalysis}
Gordon E and Okuda Y 2009 {\em Low Temperature Physics\/} {\bf 35} 209--213

\bibitem{schwarz1988three}
Schwarz K 1988 {\em Physical Review B\/} {\bf 38} 2398

\end{thebibliography}

\end{document}